\documentclass[]{article}
\begin{document}
\title{An Efficient and Robust Technique for Achieving
Self Consistency in Electronic Structure
Calculations}
\author{D.R.Bowler and M.J.Gillan\\
Department of Physics and Astronomy, 
University College London,\\
Gower Street, 
London, 
WC1E 6BT
U.K.}
\maketitle

\begin{abstract}
Pulay's Residual Metric Minimization (RMM) method is one of
the standard techniques for achieving self consistency
in {\em ab initio} electronic structure calculations.
We describe a reformulation of Pulay's RMM which guarantees
reduction of the residual at each step. The new version
avoids the use of empirical mixing parameters, and is expected to
be more robust than the original version. We present practical
tests of the new method implemented in a standard code
based on density-functional theory (DFT), pseudopotentials, and plane-wave
basis sets. The tests show improved speed in achieving
self consistency for a variety of condensed-matter systems.
\end{abstract}

\begin{center}
{\bf Accepted for publication in Chemical Physics Letters (2000)}
\end{center}

\section{Introduction}
The requirement of self consistency between the electronic charge
density and the potential plays a key role in {\em ab initio}
electronic structure calculations. The iterative search for self
consistency generally involves some form of charge density mixing at
each step. A standard and widely used mixing method is the RMM-DIIS
technique (Residual Metric Minimization -- Direct Inversion of the
Iterative Subspace), first introducted by Pulay for Hartree-Fock
calculations\cite{Pulay80,Pulay82}, but now also used in
density-functional theory\cite{Kresse95}. The modified Broyden
technique introduced by Vanderbilt and Louie\cite{Van84} and
Srivastava\cite{Sri84} and generalised by Johnson\cite{Johnson89} can
be shown\cite{Kresse96,Eyert96} to reduce to the Pulay technique for a
suitable choice of weights; in practice, these weights give the
fastest convergence, and are often used, though can lead to instability 
at convergence\cite{Johnson89,Johnson00}.  We describe here a
new technique for charge-density mixing, and we show that it is more
robust and sometimes significantly faster than the Pulay and the
modified Broyden methods.

We recall that a self-consistency cycle in an {\em ab initio}
calculation proceeds as follows: an input density $\rho ( {\bf r } )$
is used to generate a potential $v ( {\bf r} )$; from this, a Hamiltonian
or Fock matrix is formed; the eigenfunctions of the latter are
then used to create an output density $\rho^\prime ( {\bf r} )$.
The residual $R ( {\bf r} )$ associated with a given input
density is defined as $R ( {\bf r} ) \equiv \rho^\prime ( {\bf r} ) -
\rho ( {\bf r} )$. The self-consistent density is the $\rho ( {\bf r} )$
for which $R ( {\bf r} ) = 0$ everywhere. The idea of charge mixing
in its simplest form is that one should attempt to move towards
self consistency at each step by linearly `mixing' $\rho ( {\bf r} )$
and $\rho^\prime ( {\bf r} )$ in some proportions and using 
the result as input to the next cycle. The Pulay technique
is a sophisticated generalization of this idea, which works
by minimizing the norm $\mid R \mid$ of
the residual, defined as
\begin{equation}
\mid R \mid = \left[ \int d {\bf r} \, R ( {\bf r} )^2 \right]^{1/2} \; .
\end{equation}
In making a robust method for seeking self consistency, we regard it
as highly desirable that $\mid R \mid$ should decrease at every step.
We shall show that, as self consistency is approached,
our new method achieves this, though the
Pulay method does not. In the following, we summarize the Pulay
method before outlining our new formulation, which we call
guaranteed-reduction-Pulay (GR-Pulay). We then present a number of
practical test cases within DFT, which demonstrate the advantages of GR-Pulay.

\section{The GR-Pulay method}
At each iterative step, the Pulay method works with a set of
$s$ densities, which at iteration number $n$ are denoted by
$\rho_n ( {\bf r} )$, $\rho_{n-1} ( {\bf r} ) \ldots$,
$\rho_{n-s+1} ( {\bf r} )$. (In practical calculations,
$s$ is typically 5. Of course, in the early stages of the search,
when $n < s$, the set of densities is taken to be
$\rho_n ( {\bf r} ) , \ldots$ $\rho_1 ( {\bf r} )$.) 
In going to the next iteration,
a new density $\rho_{n+1} ( {\bf r} )$ is created, and the
oldest previous density $\rho_{n-s+1} ( {\bf r} )$ is discarded. The
procedure for generating $\rho_{n+1} ( {\bf r} )$ involves
the concept of the present `optimal' density $\rho_n^{\rm opt} ( {\bf r} )$,
which signifies the linear combination of the present densities:
\begin{equation}
\rho_n^{\rm opt} ( {\bf r} ) =
\sum_{i=0}^{s-1} \alpha_i \rho_{n-i} ( {\bf r} ) 
\label{eqn:comb}
\end{equation}
having the smallest norm of the residual, subject to the condition
$\sum_{i=0}^{s-1} \alpha_i =1$. To determine the coefficients $\alpha_i$,
it is assumed that we are close enough to self consistency 
for variations of the densities and their associated residuals
to be linearly related. This means that the residual associated
with any linear combination of the present densities, as in 
eqn~(\ref{eqn:comb}), is simply 
$\sum_{i=0}^{s-1} \alpha_i R_{n-i} ( {\bf r} )$, where $R_{n-i} ( {\bf r} )$
is the residual associated with input density $\rho_{n-i} ( {\bf r} )$.
The residual associated with $\rho_n^{\rm opt} ( {\bf r} )$,
denoted by $R_n^{\rm opt} ( {\bf r} )$, is then given 
by:
\begin{equation}
R_n^{\rm opt} ( {\bf r} ) =
\rho_n^{{\rm opt} \prime} ( {\bf r} ) - \rho_n^{\rm opt} ( {\bf r} ) =
\sum_{i=0}^{s-1} \alpha_i R_{n-i} ( {\bf r} ) \; ,
\end{equation}
where $\rho_n^{{\rm opt} \prime} ( {\bf r} )$ is the output charge
density associated with the input $\rho_n^{\rm opt} ( {\bf r} )$.
Since the residuals $R_{n-i} ( {\bf r} )$ are all known, the
constrained minimization of
$\mid R_n^{\rm opt} \mid$ is equivalent to the constrained minimization
of a bilinear form, and this yields a unique and simple formula
for the $\alpha_i$. 

The present optimal density $\rho_n^{\rm opt} ( {\bf r} )$ is
the `most self-consistent' linear combination of present densities.
But the new density $\rho_{n+1} ( {\bf r} )$ clearly
cannot be chosen to be $\rho_n^{\rm opt} ( {\bf r} )$. The reason
for this is that each new iteration must introduce new variations
of the density, so that $\rho_{n+1} ( {\bf r} )$ cannot be allowed to lie
in the subspace spanned by the present densities. In the
conventional implementation of the Pulay scheme, the new density is usually
chosen to be a linear combination of $\rho_n^{\rm opt} ( {\bf r} )$
and its output density $\rho_n^{{\rm opt} \prime} ( {\bf r} )$:
\begin{equation}
\rho_{n+1} ( {\bf r} ) =
( 1 - A ) \rho_n^{\rm opt} ( {\bf r} ) + 
A \rho_n^{{\rm opt} \prime} ( {\bf r} ) \; .
\end{equation}
The value of the mixing coefficient $A$ is empirically chosen
(typically to be about 0.8), and the efficiency of the
self-consistency search depends on the choice of $A$. If $A$ is not
appropriately chosen, a variety of problems can occur, including slow
convergence or even failure to converge. The problem is that the best
choice of $A$ depends on the physical system being treated\footnote{We note
that Pulay\protect\cite{Pulay80,Pulay82} implicitly chose A=1; we also 
note that as the scheme enters the linear regime, the value of A becomes
unimportant.}.

The new mixing scheme proposed here resembles the Pulay method
in that it works with $s$ densities $\rho_n ( {\bf r} )$,
$\rho_{n-1} ( {\bf r} ) , \ldots$, $\rho_{n-s+1} ( {\bf r} )$
at each step, and goes from one iteration to the next by discarding
the oldest density $\rho_{n-s+1} ( {\bf r} )$ and creating
a new density $\rho_{n+1} ( {\bf r} )$. However, the set of $s$
densities is required to have a crucial property:
the norm of the
residual $R_n ( {\bf r} )$ associated with $\rho_n ( {\bf r} )$ is required
to be no greater than the norm of the residual associated with
any linear combination $\sum_{i=0}^{s-1} \alpha_i \rho_{n-i} ( {\bf r} )$,
with the usual condition $\sum_{i=0}^{s-1} \alpha_i = 1$.
We shall see immediately how to ensure this property. Our procedure
for generating $\rho_{n+1} ( {\bf r} )$ is as follows:

\begin{itemize}
\item
delete $\rho_{n-s+1}$ and add $\rho_n^\prime$, so that the set
of densities is $\rho_n^\prime$, $\rho_n$, $\rho_{n-1} , \ldots$
$\rho_{n-s+2}$.

\item
put $\rho_n^\prime$ through a cycle, so that we have its output
$\rho_n^{\prime \prime}$
and hence a full set of residuals $R_n^\prime$, $R_n$, $R_{n-1} \ldots$
$R_{n-s+2}$.

\item
make linear combinations:
\begin{equation}
{\hat{\rho}}_n = \alpha_1 \rho_n^\prime +
\alpha_2 \rho_n + \alpha_3 \rho_{n-1} + \ldots
+ \alpha_s \rho_{n-s+2} \; ,
\end{equation}
and determine the coefficients $\alpha_i$ so as to minimize the
norm of the residual of ${\hat{\rho}}_n$,
subject to the usual condition that the $\alpha_i$ sum
to unity. Denote by $\rho_{n+1}$
the density ${\hat{\rho}}_n$ that yields the minimum residual.

\item
delete $\rho_n^\prime$ and replace by $\rho_{n+1}$, so that the new
set of densities is $\rho_{n+1}$, $\rho_n , \ldots$ $\rho_{n-s+2}$.
\end{itemize}
Clearly the new set has the same property as the old: the newest density
$\rho_{n+1}$ has the minimal residual norm that can be achieved
by taking linear combinations of the present densities.
In particular, $\mid R_{n+1} \mid \, < \, \mid R_n \mid$, so that
the residual at iteration $n+1$ is less than that at iteration $n$.
Moreover, it can be shown that the decrease of the residual in
an $s$-level scheme is greater than the decrease in an $s^\prime$-level
scheme, provided $s > s^\prime$. We note that in GR-Pulay, as in
the original Pulay scheme, the computational effort in each iteration 
consists of a single self-consistent cycle. This is true, provided
$R_{n+1} ( {\bf r} )$ is accurately enough given by the linear
approximation, which is clearly the case as self-consistency is
approached.

\section{Applications}
Although our GR-Pulay scheme is completely general, our practical
interest here is in the pseudopotential, plane-wave implementation of
DFT, and our tests have been done on a variety of condensed-matter
systems using the CASTEP code\cite{Payne92}. The use of a modified
Broyden technique\cite{Johnson89} (shown by Kresse\cite{Kresse96} and
Eyert\cite{Eyert96} to reduce to the usual Pulay technique for a
suitable choice of the weights) in DFT condensed-matter work has been
explored by many workers, including Johnson and Kresse, and it is one
of the standard options in CASTEP; it is implemented with the weights
chosen so that it is directly equivalent to the Pulay technique. The
systems chosen for our tests are: bulk silicon (two atom primitive
cell); bulk magnesium chloride (three atom cell); bulk aluminium (one
atom stacked fcc); bulk plutonium dioxide (three atoms stacked fcc);
bulk iron (one atom stacked fcc); and the platinum(001) surface (five
atoms in five layers, slab geometry).  These systems provide example
of insulating, semiconducting and metallic behaviour, as well as
extreme inhomogeneity of electron density.

We have made efforts to achieve an unbiased comparison of Broyden and
GR-Pulay; all parameters in each run were the same for the two
techniques, as were the initial conditions.  When searching for
self-consistency, CASTEP applied a criterion of energy change between
successive iterations to the Broyden method, while we apply a
criterion of absolute size of the norm of the residual (expressed as a
fraction of the norm of the charge density) to the GR-Pulay method. We
checked that in all cases our criterion was as strict as the energy
difference criterion (i.e. the norm of the residual when CASTEP had
converged using Broyden was no smaller than the norm of the residual
when CASTEP had converged using GR-Pulay).  We note that, in the
context of the search for self-consistency between a charge density
and a potential, it is important to apply a convergence criterion to
the norm of the residual (for this is what determines whether or not
self-consistency has been reached) and not an energy difference; cases
where the change in energy is small from one iteration to the next,
but the norm of the residual is relatively large, can be envisaged.

In Table~\ref{tab:timings} we report timings for the Pulay/Broyden
technique and our GR-Pulay technique within CASTEP, run on a PC
(400MHz Pentium II with 256MB of memory) under Linux.  The
calculations employed standard pseudopotentials (either
norm-conserving\cite{Kerker,KB} or ultrasoft\cite{Vand}) and plane
wave cutoffs of between 100 and 300 eV (depending on the system; the
efficacy of individual methods should be independent of the plane wave
cutoff).  The criteria for self consistency are an energy change of
5$\times 10^{-6}$eV per atom for the Pulay/Broyden technique and
$|R|/|\rho| < 10^{-4}$ for the GR-Pulay technique.  The first two runs
(Si and MgCl$_2$) involved relaxation of the atomic positions and the
third (Al) the determination of the unit cell size (these were pursued
until the RMS force on the atoms was below 0.01 eV/\AA, and the RMS
stress below 0.1 GPa).  The next three cases (PuO$_2$, Fe and Pt(001))
were all single point energy calculations.  For the first three, the
number of iterations required to reach the self-consistent ground
state starting from scratch (i.e. at the first iteration) is given in
brackets after the time.  For the last three, the total number of
self-consistent iterations is given in brackets after the time.  Our
timings demonstrate that GR-Pulay can be over twice as fast as the
Pulay/Broyden method; in all cases, this is because GR-Pulay requires
fewer iterations to find the ground state than Pulay/Broyden.  It is
also related to the number of line searches performed during wave
function minimisation between self-consistent iterations; modern plane
wave DFT codes perform several line searches (until either convergence
or a maximum number of iterations are reached), and differing numbers
of these searches will also contribute to the different times.
Monitoring the rate of decrease of the residuals in both cases showed
that the GR-Pulay method typically achieved a much greater rate than
Pulay/Broyden (which is not surprising, given the results).

\section{Conclusions}
We have shown that our proposed reformulation of the Pulay
mixing scheme offers three kinds of advantage over the original
scheme. First, no empirically chosen mixing parameters are needed,
so that less experience is required to use the scheme successfully.
Second, the residual is guaranteed to decrease at every iteration, so that
instability cannot occur, at least as self-consistency is approached.
Third, our practical tests show that the new scheme can sometimes
be significantly faster than the original one.

\section*{Acknowledgements}

We are grateful to Professor D.D.Johnson and Professor V.Eyert for
useful discussions.  DRB acknowledges the support of the EPSRC through
a postdoctoral fellowship in theoretical physics (Grant GR/M71640).

\begin{table}[h]
\begin{center}
\begin{tabular}{|l|l|l|}
\hline
System   & Pulay/Broyden & GR-Pulay \\
\hline
Si(R)    &   6.93 (9) &   5.56 (7)  \\
MgCl$_2$(R)  &  30.19 (13) &  25.35 (13)  \\
Al(CR)   &  29.50 (36) &  12.34 (12)  \\
\hline
PuO$_2$  &  20.91 (15) &  16.47 (10)  \\
Fe(spin) &  63.02 (34) &  33.05 (14)  \\
Pt(001)  & 670.47 (15) & 688.72 (14)  \\
\hline
\end{tabular}
\end{center}
\caption{Timings (in seconds) for the Pulay/Broyden and GR-Pulay methods
applied to different systems using the CASTEP code.  Numbers in
brackets after the time indicate iterations required to reach the
self-consistent ground state starting from initial conditions for the
first iteration.  (R) indicates geometry optimisation, (CR) cell
optimisation and no letter a single point energy.}
\label{tab:timings}
\end{table}

\end{document}